\documentstyle[twocolumn,aps,pre,epsfig,amsfonts]{revtex} 
\begin{document}                                     
\title{Generalized Force Model of Traffic Dynamics}
\author{Dirk Helbing and Benno Tilch}
\address{II. Institute of Theoretical Physics, University of
Stuttgart, 70550 Stuttgart, Germany\\
{\tt http://www.theo2.physik.uni-stuttgart.de/helbing.html}}
\maketitle      
\draft
\begin{abstract}
  Floating car data of car-following behavior in cities were compared to
  existing microsimulation models, after their parameters had been 
  calibrated to the experimental data. With these parameter values,
  additional simulations have been carried out, e.g.~of a moving car which
  approaches a stopped car. It turned out
  that, in order to manage such kinds of situations without producing
  accidents, improved traffic models are needed. Good results have been
  obtained with the proposed generalized force model. 
\end{abstract}
%
%
\section{Introduction}    

During the last five years, 
gas-kinetic \cite{Hab,kin}, fluid-dynamic \cite{Hab,kin,Macro}, and
other models have been developed,
aiming at an understanding of
stop-and-go traffic. 
The topic is related to the fields 
of non-linear dynamics \cite{Macro}, 
phase transitions \cite{phase}, and
stochastic processes \cite{Cell1}. In addition, microscopic traffic models
were proposed for the description of interacting driver-vehicle units.
They can be classified into cellular automata
models \cite{Cell1}, which are discrete in space and time, 
and continuous models \cite{Hab,FL,OVM}.
The latter are required for detail studies of car-following behavior and
traffic instabilities, which are necessary for an investigation
of the consequences of technical optimization measures (e.g.~of
autopilots for an automatic control of vehicle acceleration and braking).
\par
Therefore, a research group of the Bosch GmbH has
recently recorded follow-the-leader data by means of a floating car
$\alpha$ which measured the vehicle speed $v_\alpha$, the netto
distance $s_\alpha$ to the car in front, the acceleration $a_\alpha$ of it,
and the relative velocity $\Delta v_\alpha$. 
By a correlation analysis it was demonstrated that, 
among all possible combinations of
subsets of these four quantities, $s_\alpha$, 
$\Delta v_\alpha$, and $v_\alpha$ are the most
significant variables for the description of vehicle dynamics \cite{Bosch}.
In Sec.~\ref{genforce}, we will find plausible reasons for this.
\par
\begin{figure}
\begin{center}
\includegraphics[width=8cm]{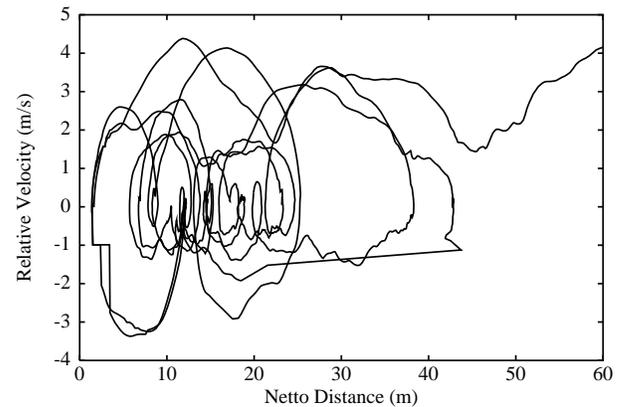}
\end{center}
\caption[]{The follow-the-leader data show 
    the oscillatory nature of the relative 
    motion of vehicles.\label{F6}}
\end{figure}

The follow-the-leader data, if plotted in the 
$s_\alpha$-$\Delta v_\alpha$-plane, 
show the characteristic oscillation of vehicle motion around
states with relative velocity zero (cf.~Fig.~\ref{F6}),
which was already reported by Hoefs \cite{Hoefs}. Except for
the previously mentioned significance analysis, the data were also
used for calibrating existing microsimulation models.
With the resulting optimal sets of parameter values, the models
were simulated for the observed situation. That is, the first vehicle was
moved according to its measured velocity,
and the following vehicle was simulated according to the respective model
under consideration, starting with the the same initial velocity and distance
like the floating car. The average relative quadratic  
deviation $D$ between the simulated and actually
measured distance  (cf.~Eq.~(\ref{De}))
was used as a measure for the goodness of fit of the respective model
\cite{Bosch}.
\par
In order to get improved results, 
we have developed a generalized
force model, in which each term and each parameter has a clear meaning.
Moreover, by parameter calibration it turns out that all model parameters
have the right order of magnitude. Therefore, it can be easily said,
how the parameter values will look like, if the speed limit,
the acceleration capability, the average vehicle length, the visibility,
or the reaction time is modified (e.g.~due to technical measures). 
In addition, this model reaches a better goodness of fit at a reduced number
of model parameters than the previous models. 
Finally, the generalized force model manages to cope successfully
with particular situations like vehicles approaching standing cars,
in which other models produce accidents.

\section{Discussion of previous models}

The first microscopic traffic models were developed in the 1960ies.
Many of them are special cases of the follow-the-leader model
proposed by Gazis, Herman, and Rothery \cite{FL}. This assumes that
the dynamics of a vehicle $\alpha$ with velocity $v_\alpha(t)$ at
place $x_\alpha(t)$ is given by the equation of motion
\begin{equation}
 \frac{dx_\alpha(t)}{dt} = v_\alpha(t)
\end{equation}
and the acceleration equation
\begin{equation}
 \frac{dv_\alpha(t+T)}{dt} = \kappa(t+T) [v_{\alpha-1}(t) - v_\alpha(t)] \, .
\end{equation}
According to this, a driver adapts to the velocity 
$v_{\alpha-1}(t)$ of the car in front, but this is delayed by the
adaptation time $T\approx 1.3$\,s. The deceleration is proportional
to the relative velocity 
\begin{equation}
  \Delta v_\alpha = v_{\alpha} - v_{\alpha -1} \, ,
\end{equation} 
where the proportionality factor $\kappa$ reflects the sensitivity to 
the stimulus $\Delta v_\alpha$. The sensitivity was assumed
to depend on the vehicle velocity and on the brutto distance 
\begin{equation}
 S_\alpha = x_{\alpha-1} - x_\alpha 
\end{equation}
in the following way:
\begin{equation}
 \kappa(t+T) = \kappa_0 \frac{[v_{\alpha}(t+T)]^m}{[S_\alpha(t)]^l} \, .
\end{equation}
This choice allowed to fit all equilibrium 
velocity-density relations of the form
\begin{equation}
 V_{\rm e}(\rho) = v_0 \left[ 1 - \left(\frac{\rho}{\rho_{\rm max}}
   \right)^{l-1} \right]^{1/(1-m)} 
\end{equation}
by appropriate specification of the exponents $l$ and $m$
($v_0$ = maximum velocity, $\rho$ = spatial vehicle density,
$\rho_{\rm max}$ = maximum vehicle density). The best fit was
reached for fractional exponents
$m\approx 0.8$ and $l \approx 2.8$, so that this model has
no obvious interpretation. Apart from that, the model does not
allow to distinguish drivers with different preferred velocity,
and it cannot describe the acceleration of a single vehicle correctly.
\par
Only a few years ago, Bando, Sugiyama {\em et al.} proposed a very
charming microscopic traffic model. 
Despite its simplicity and its few parameters,  
their optimal velocity model (OVM) describes many properties 
of real traffic flows \cite{OVM} 
and is easily interpretable. It is based on the acceleration
equation
\begin{equation}
 \frac{dv_\alpha(t)}{dt} = \kappa [ V(s_\alpha) - v_\alpha(t) ] \, ,
\end{equation} 
so that the vehicles adapt to a distance-dependent optimal velocity
\begin{equation}
 V(s_\alpha) = V_1 + V_2 \tanh (C_1 s_\alpha  - C_2)
\label{Ve}
\end{equation}
with a certain relaxation time $\tau = 1/\kappa$. Here,
\begin{equation}
 s_\alpha = x_{\alpha-1} - x_\alpha - l_{\alpha-1}
 = S_\alpha - l_{\alpha -1}
\end{equation}
denotes the netto distance, where $l_\alpha$ means the length of
vehicle $\alpha$. Like the
follow-the-leader models, the optimal velocity model 
is able to describe the formation
of stop-and-go waves and emergent traffic jams, but it overcomes
the afore mentioned problems.  
\par
We carried out a calibration of the optimal velocity model with respect to
the empirical follow-the-leader data, which we obtained from Bosch.
The optimization procedure based on the evolutionary
Boltzmann strategy \cite{Opt}, 
and the optimization criterion was the average relative quadratic  
deviation 
\begin{equation}
  D = \frac{1}{N} \sum_{t=1}^N \left( \frac{s_\alpha(t) - s_\alpha^{\rm m}(t)}
    {s_\alpha^{\rm m}(t)} \right)^2
\label{De}
\end{equation}
of the simulated distance $s_\alpha(t)$ from the measured vehicle distance
$s_\alpha^{\rm m}(t)$. The resulting 
optimal parameter values for city traffic in Stuttgart are
$\kappa = 0.85\,\mbox{s}^{-1}$, $V_1 = 6.75$\,m/s, $V_2 = 7.91$\,m/s,
$C_1 = 0.13\,\mbox{m}^{-1}$, and $C_2 = 1.57$.
\par
\begin{figure}
  \begin{center}
    \includegraphics[width=8cm]{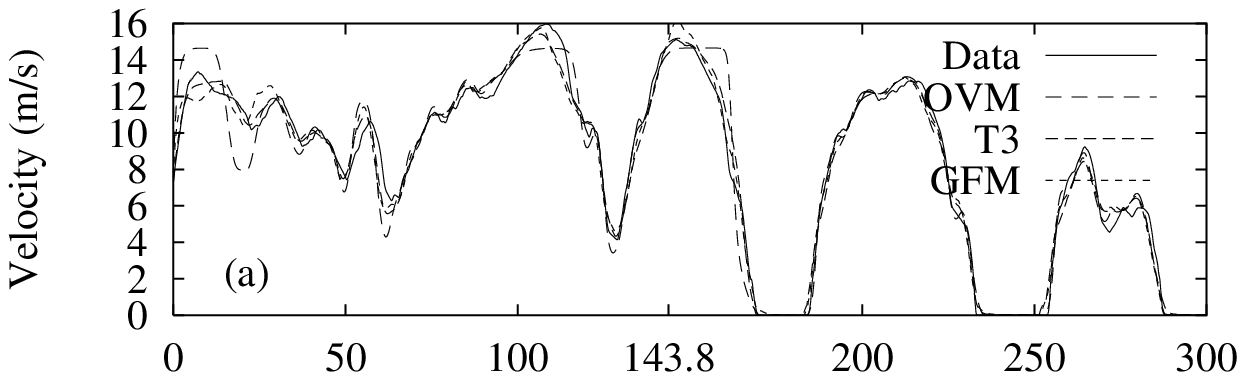}
    \includegraphics[width=8cm]{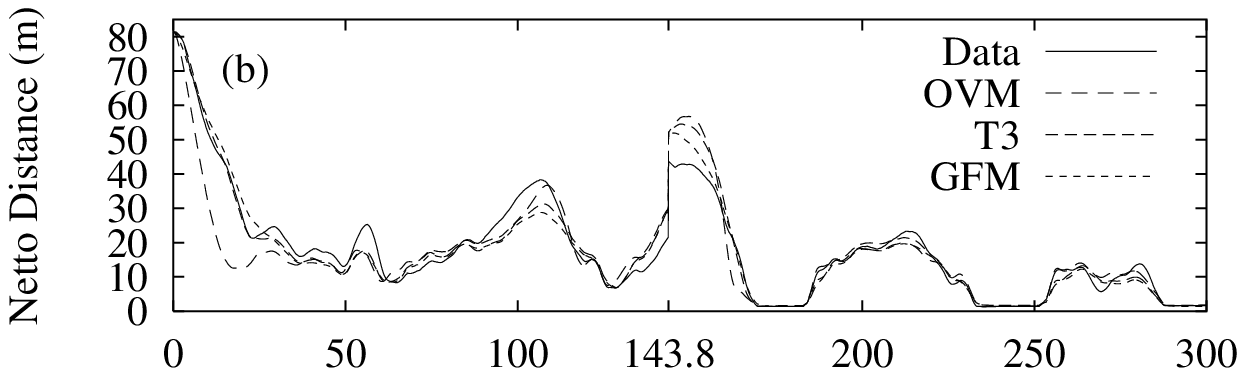}
    \includegraphics[width=8cm]{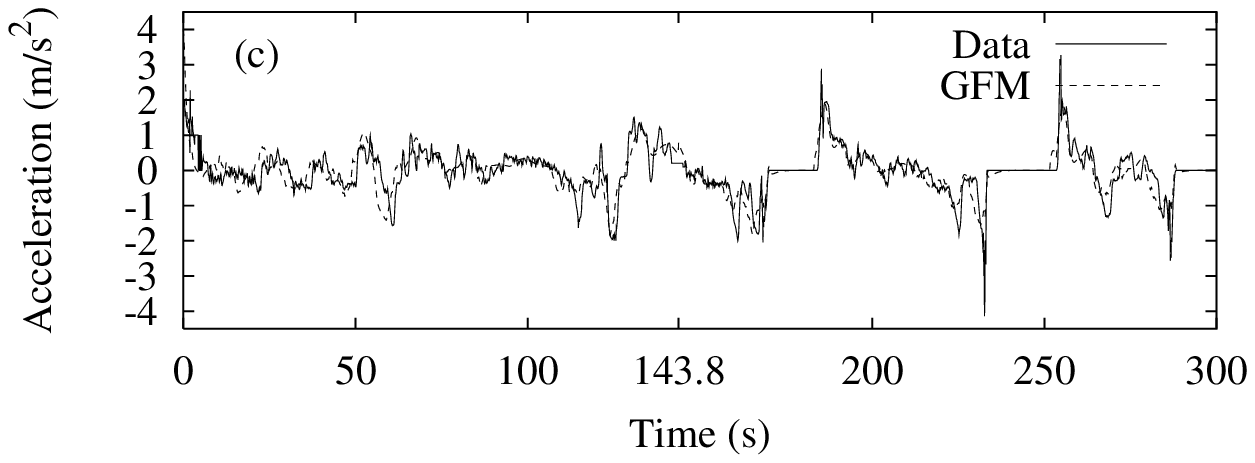}
    \caption[]{The time-dependent velocity $v_\alpha(t)$ (a), 
      distance $s_\alpha(t)$ (b), and acceleration $dv_\alpha(t)/dt$ (c)
      according to the optimal velocity model
      (OVM), the general force model (GFM), and the T3 model in comparison with
      follow-the-leader data of city traffic. According to (a), most of the 
      models compare well to the measured velocities. Only the OVM shows a
      significant overshooting, indicating too large accelerations. Fitting
      of the vehicle distances is a much harder task, as shown in (b), but the
      T3 model and the GFM perform well. In (c), one can see that the
      empirical accelerations and decelerations are usually limited to 
      the range between $-3$\,m/s$^2$ and $+4$\,m/s$^2$, which is
      met by the GFM.\\ Note that the vehicles had to stop
      three times due to red traffic lights (during the periods between
      169.9 and 184 seconds, 233.5 and 253.5 seconds, 
      and after $t=288$\,s). At time $t=143.8$\,s, the
      vehicle in front was turning right, so that the second car was
      following another vehicle, afterwards.\label{F5}}
    \end{center}
\end{figure}
\begin{figure}
  \begin{center}
    \includegraphics[width=8cm]{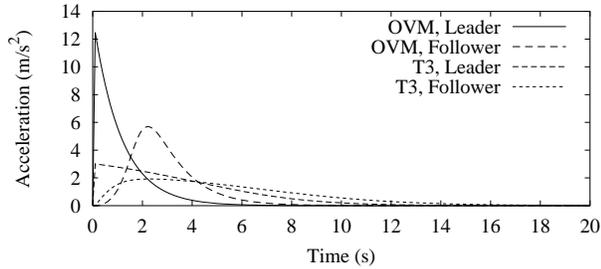}
    \caption[]{Acceleration of an unobstructed vehicle  and of
      a following vehicle according to the optimal velocity model (OVM)
      and the T3 model. Initially, both vehicles are at rest.\label{F2}}
  \end{center}
\end{figure}
\begin{figure}
  \begin{center}
    \includegraphics[width=8cm]{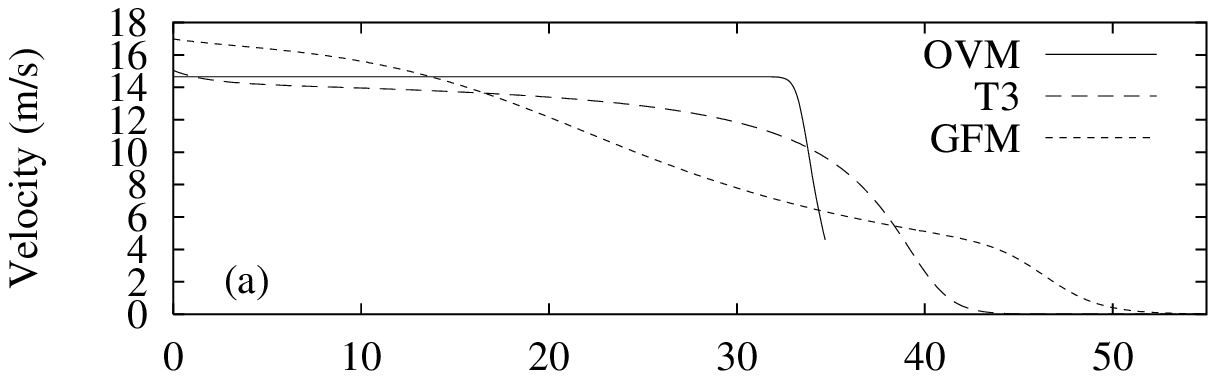}
    \includegraphics[width=8cm]{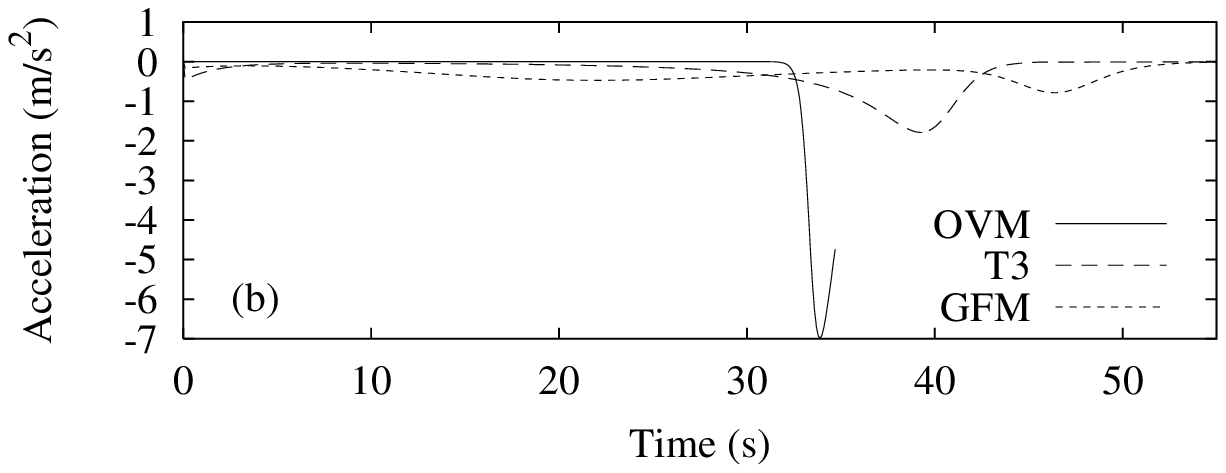}
    \caption[]{Time-dependent velocity $v_\alpha(t)$ (a) and acceleration
      $dv_\alpha(t)/dt$ (b) for a vehicle which approaches a standing vehicle
      according to the different discussed simulation models. 
      The optimal velocity model
      produces an accident at time $t=34.7$\,s.\label{F3}}
    \end{center}
\end{figure}

A comparison with the data 
shows that the extremely short relaxation time $\tau = 1/\kappa = 1.17$\,s 
results in too high values of acceleration, which leads to an overshooting 
of the vehicle velocity (cf.~Fig.~\ref{F5}a). 
The unrealistically high accelerations also become obvious in
Figure~\ref{F2}, since empirical accelerations are limited to
4\,m/s$^2$ (cf.~Fig.~\ref{F5}c). 
A similar problem occurs with the deceleration behavior,
if a standing car (e.g.~at the end of a traffic jam or in front of
a red traffic light) is approached from a large distance 
by an initially freely moving car. It turns out that the moving vehicle
reacts too late to the vehicle at rest. The values of deceleration are
unrealistic large, but still not sufficient to avoid an accident 
(cf.~Fig.~\ref{F3}).
\par
These problems are solved by the {\em T3 model} 
\begin{equation}
 \frac{dv_\alpha}{dt} = \frac{1+b_1 v_\alpha + b_2 s_\alpha + b_3 v_\alpha
   s_\alpha + b_4 v_{\alpha-1} + b_5 v_\alpha v_{\alpha-1}}
 {c_0 + c_1v_\alpha + c_2 s_\alpha + c_3 v_\alpha s_\alpha
 + c_4 v_{\alpha-1} + c_5 v_\alpha v_{\alpha-1}} 
\label{reg} 
\end{equation} 
proposed by Bosch \cite{Bosch}. $b_k$ and $c_k$ are model parameters. 
The regression model (\ref{reg}), which is based on a rational function,
describes all aspects of vehicle dynamics in cities 
realistically (cf.~Figs.~\ref{F5}--\ref{F3}), but at the
cost of additional parameters. Whereas the optimal velocity model
needs only 5 model parameters, the T3 model contains 11 parameters.
If the model equations are scaled to dimensionless equations (by scaling
space or velocity and time by characteristic model quantities), the number of
parameters is reduced by 2. 
\par
In the following section, we will propose an alternative model
which reaches about the same goodness of fit as the T3 model, but
with a considerably smaller number of parameters.  
 
\section{The generalized force model} \label{genforce}

Motivated by the success of so-called social force models 
in the description of behavioral changes \cite{SF,Hab}, 
especially of pedestrian 
dynamics \cite{Fuss,Hab}, we developed a related model for 
the dynamics of interacting vehicles. In setting up an
equation of motion by specifying the effective acceleration and deceleration
forces, the approach is analogous to the 
molecular-dynamic method which is used for the simulation of many-particle
systems \cite{MD}, e.g.~of driven granular media \cite{gran}.
\par
Besides of various methodological similarities in the theoretical treatment
of traffic and granular flows, there are also phenomenological analogies
like the formation of density waves \cite{analog}. In both cases,
the interactions are dissipative, i.e. they do not conserve kinetic energy.
However, there are also differences. For granular media, the interaction
forces are short-ranged and belong to collision processes, 
where particles touch and
temporarily deform each other. Vehicle interactions are long-ranged and
correspond to deceleration maneuvers. They are usually not related to 
collisions (i.e.~accidents), since the drivers try to keep a
safe distance $s$ from each other which can
be considerably larger than the vehicle length
(cf.~Eq.~(\ref{safety})). Therefore, the effective
space requirements of vehicles are much larger than their actual size.
Moreover, vehicular interactions do not conserve momentum 
(in contrast to granular ones).
\par
Another difference between granular and traffic dynamics is, that
the laws of granular interactions are very well known, pretty much like
the basic laws of physics, whereas the laws of driver-vehicle dynamics 
(if they exist at all) are still to be established. The particular challenge 
of modeling vehicle dynamics is its dependence on factors like perceptions,
psychological motivations and reactions, or social behaviors. 
Thus, in contrast to physical processes, driver behavior cannot be expected 
to be describeable by a few natural constants.
\par
According to the social force concept, the amount and direction of a
behavioral change
(here: the temporal change of velocity, i.e.~the acceleration) is given
by a sum of generalized forces. These reflect the different motivations
which an individual feels at the same time, e.g.~in response to their
respective environment. Since these forces do not fulfill Newton's
laws like {\em actio = reactio}, they are called {\em generalized
forces.} Alternatively, they are named {\em behavioral} or
{\em social forces}, since they mostly correspond to social interactions. 
The success of this approach in describing traffic dynamics 
is based on the fact that driver
reactions to typical traffic situations are more or less 
automatic and determined by the optimal behavioral strategy (which
is the results of an initial learning process). A detailled motivation,
description, and discussion of the social force concept is given in
Refs.~\cite{SF,Hab,Fuss}.
\par
The driver behavior is mainly given by the motivation to reach a
certain desired velocity $v_\alpha^0$ (which will be reflected by an
acceleration force $f_\alpha^0$) and by the motivation to keep a
safe distance to other cars $\beta$ (which will be described by repulsive
interaction forces $f_{\alpha,\beta}$):
\begin{equation}
 \frac{dv_\alpha}{dt} = f_\alpha^0(v_\alpha) +
 \sum_{\beta (\ne \alpha)} f_{\alpha,\beta}(x_\alpha, v_\alpha;
 x_\beta, v_\beta) + \xi_\alpha(t) \, .
\end{equation}
The fluctuating force $\xi_\alpha(t)$ may be used to include 
individual variations of driver behavior, but in our present investigations
it was set to zero. If we assume that the acceleration force is proportional
to the difference between desired and actual velocity 
and suppose that the most important interaction
concerns the car $(\alpha-1)$ in front, we end up with
\begin{equation}
 \frac{dv_\alpha}{dt} = \frac{v_\alpha^0 - v_\alpha}{\tau_\alpha} +
 f_{\alpha,\alpha-1}(x_\alpha, v_\alpha;
 x_{\alpha-1}, v_{\alpha-1}) \, .
\label{force}
\end{equation}
The acceleration time $\tau_\alpha$ is a third of the time which a freely
accelerating vehicle needs to reach 95\% of the desired
velocity. 
\par
Now, we have to specify the interaction force $f_{\alpha,\alpha-1}$.
For 
\begin{equation}
 f_{\alpha,\alpha-1} = \frac{V(s_\alpha) - v_\alpha^0}{\tau_\alpha}
\end{equation}
and $\tau_\alpha = 1/\kappa$, we would obtain the optimal velocity model,
again. We extend this relation by a complementary
term which should guarantee early enough and sufficient braking
in cases of large relative velocities $\Delta v_\alpha$. This term should
increase with growing velocity difference $\Delta v_\alpha$, but it
should be only effective, if the velocity of the following vehicle is
larger than that of the leading vehicle, i.e.~if the Heaviside function
$\Theta(\Delta v_\alpha)$ is equal to 1. Moreover, the additional 
deceleration term should increase with decreasing distance $s_\alpha$,
but vanish for large distance $s_\alpha \rightarrow \infty$. 
The braking time $\tau'_\alpha$ belonging to this term should
be smaller than $\tau_\alpha$, since deceleration capabilities of vehicles are
greater than acceleration capabilities. We chose the following formula which
meets the above conditions:
\begin{equation}
 f_{\alpha,\alpha-1} = \frac{V(s_\alpha) - v_\alpha^0}{\tau_\alpha}
 - \frac{\Delta v_\alpha \Theta(\Delta v_\alpha)}{\tau'_\alpha}
 \;\mbox{e}^{ - [s_\alpha - s(v_\alpha)]/R'_\alpha} \, .
\label{interact}
\end{equation}
This formula takes into account that vehicles prefer to keep a certain
velocity-dependent safe distance 
\begin{equation}
 s(v_\alpha) = d_\alpha + T_\alpha v_\alpha \, ,
\label{safety}
\end{equation}
where $d_\alpha$ is the minimal vehicle distance and $T_\alpha$ is the
safe time headway (i.e.~about the reaction time). $R'_\alpha$ can be 
interpreted as range of the braking interaction.
\par
\begin{figure}
  \begin{center}
    \includegraphics[width=8cm]{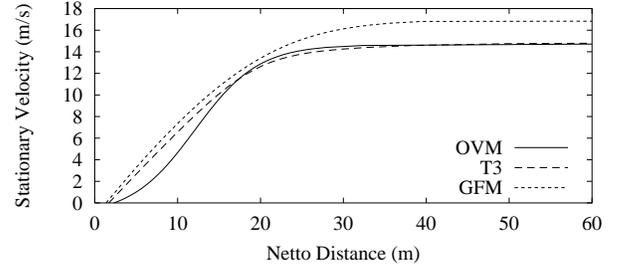}
    \caption[]{Velocity-distance relations of the 
      different traffic models in the stationary case, if
      all vehicles have identical parameters.\label{F4}} 
  \end{center}
\end{figure}

We can further reduce the number of parameters (and the numerical effort), 
if we replace the previous $V(s_\alpha)$-function (\ref{Ve}) by
\begin{equation}
 V_\alpha(s_\alpha,v_\alpha) = v_\alpha^0 \big\{1 - \mbox{e}^{ 
 - [s_\alpha - s(v_\alpha)]/
 R_\alpha} \big\}  \, .  
\label{vspec}
\end{equation} 
In case of identical model parameters of all vehicles, the corresponding
equilibrium velocity-distance relation results from 
the implicit condition $v_\alpha = V_\alpha(s_\alpha,v_\alpha)$ 
and is depicted in Figure~\ref{F4}.
\par
The traffic model defined by equations (\ref{force}), 
(\ref{interact}), and (\ref{vspec}) will, in the following, be called the 
{\em generalized force model} of traffic dynamics (GFM). 
Since all its seven parameters have a clear and measurable meaning,
they should have the right order of magnitude.
A calibration with respect to the follow-the-leader data shows that this is
indeed the case. We found the following optimal parameter values:
$v_\alpha^0 = 16.98$\,m/s, $\tau_\alpha = 2.45$\,s, 
$d_\alpha = 1.38$\,m, $T_\alpha = 0.74$\,s, 
$\tau'_\alpha = 0.77$\,s, 
$R_\alpha = 5.59$\,m, and $R'_\alpha = 98.78$\,m.
Now, the acceleration time $\tau_\alpha$ is more than twice as large as in the
optimal velocity model, the braking time $\tau'_\alpha$ is smaller than
$\tau_\alpha$, as demanded, and the reaction time $T_\alpha$ is also
realistic. Note 
that the range $R_\alpha$ of the acceleration interaction is much shorter
than the range $R'_\alpha$ of the deceleration interaction. This is
not only sensible, it is also the reason for the astonishingly good
agreement with the empirical data. Table~\ref{table1} compares the
minimal values of the average relative quadratic deviation $D$ 
that could be reached for the different discussed traffic models 
by evolutionary parameter optimization \cite{Opt}. 
(The advantage of the applied Boltzmann strategy is that this particular
gradient method escapes local minima by means of a fluctuation term with
eventually decreasing variance.)
\par
It turns out that the optimal velocity model is considerably
improved by the T3 model. This is not surprising, since the goodness of fit
should increase with the number of model parameters.
Nevertheless, the generalized force model reaches the best agreement 
with the data, although it includes only two third 
of the number of parameters of the T3 model.
The simulation results for the generalized force model
are depicted in Figure~\ref{F5}. 
Finally, the representation of the relative vehicle movement
in the $\Delta v_\alpha$-$s_\alpha$-plane
shows the expected oscillatory character of follow-the-leader behavior, which
can cause the development of stop-and-go traffic (cf. Fig.~\ref{F7}).
\par
In comparison with physical models, the generalized force model still
seems to contain a lot of parameters. However, let us discuss this in more
detail for the previously mentioned molecular-dynamic models of granular media.
If we want to describe interactions of smooth, inelastic, and spherical grains,
only, we need to know the particle size and the normal restitution
coefficient related to translational energy dissipation. In cases
of rough spheres, we require two additional parameters,
a tangential restitution coefficient and a friction coefficient
\cite{coll}.
Moreover, if the grains are non-spherical, the situation becomes
even more complicated. In conclusion, the small number of parameters
occuring in physical models are often a result of simplifications and
idealizations.
\par
Finally, let us check the plausibility of the generalized force model.
In order to do this, we rewrite the model in the form
\begin{equation}
 \frac{dv_\alpha}{dt} = \frac{V_\alpha^*(s_\alpha,v_\alpha,\Delta v_\alpha) - 
 v_\alpha(t)}{\tau_\alpha^*}
\end{equation}
with
\begin{equation}
 \frac{1}{\tau_\alpha^*} = \frac{1}{\tau_\alpha} +
 \frac{\Theta(\Delta v_\alpha)}{\tau_\alpha^{\prime\prime}} 
\end{equation}
and
\begin{equation}
 V_\alpha^* = \frac{\tau_\alpha^{\prime\prime} V_\alpha + \tau_\alpha
 \Theta(\Delta v_\alpha) v_{\alpha-1}}{\tau_\alpha^{\prime\prime} 
 + \tau_\alpha \Theta(\Delta v_\alpha)} \, , 
\end{equation}
where
\begin{equation}
 \tau_\alpha^{\prime\prime} = \tau'_\alpha \exp \{ [s_\alpha -
     s(v_\alpha)]/R'_\alpha\} \, . 
\end{equation}
For small velocity differences $\Delta v_\alpha$ or large distances $s_\alpha$,
we find
\begin{equation}
 \frac{dv_\alpha(t)}{dt} \approx \frac{V_\alpha(s_\alpha,v_\alpha) 
 - v_\alpha(t)}{\tau_\alpha} \, ,
\end{equation}
so that vehicles try to approach the optimal velocity $V_\alpha$ 
with a relaxation time $\tau_\alpha$. This corresponds to the
optimal velocity model, but with a velocity-dependent function
$V_\alpha(s_\alpha,v_\alpha)$. If a vehicle is faster than the
leading vehicle (i.e.~$\Delta v_\alpha > 0$) and its distance is
sufficiently small, we have
\begin{equation}
 \frac{dv_\alpha(t)}{dt} \approx \frac{v_{\alpha-1}(t) 
 - v_\alpha(t)}{\tau_\alpha^{\prime\prime}} \, ,
\end{equation}
which coincides with a car-following model in which
the following vehicle adapts to the velocity of the
vehicle in front. According to the formula for $\tau_\alpha^{\prime\prime}$,
the deceleration is the stronger, the closer the vehicles come to each other.  
Therefore, the limiting cases of the generalized force model behave
very reasonably.

\section{Summary and discussion}

We have calibrated several microscopic traffic models to city traffic,
compared them with empirical follow-the-leader data, and investigated 
their properties. It turned out that one model showed too large accelerations
and decelerations, but nevertheless caused accidents in certain situations.
Another model was a regression model, so that the meaning of the model
and its parameters was not clear. Therefore, we developed the
generalized force model, which reached the best agreement with
the empirical data, although it had only two parameters more than the
optimal velocity model and four parameters less than the T3 model.
Another advantage of the generalized force model is that all its parameters
are easily interpretable and have the expected order of magnitude.
Therefore, it can be immediately said, how the parameters will differ
between fast cars, slow cars, and trucks (the latter being characterized by
small $v_\alpha^0$, but large $\tau_\alpha$ and $\tau'_\alpha$).
It can also be predicted what happens, if the speed limit (i.e.~$v_\alpha^0$)
is changed, if the weather conditions are bad (greater $\tau'_\alpha$,
but smaller $v_\alpha^0$ and $R'_\alpha$), if roads would be used by vehicles
with smaller length $l_\alpha$, if the reaction time $T_\alpha$ could
be reduced (by means of technical measures like an autopilot), etc.
For these reasons, the generalized force model is an ideal tool for 
carrying out detail studies of traffic flow, as well as for
developing and testing traffic optimization measures.  
\par
The simulation of large vehicle numbers is completely analogous to
molecular-dynamic simulation studies of many-particle systems, 
e.g.~of granular flows.
The parameters of the different driver-vehicle units are specified
individually ($\alpha$-dependent), then. 
In this case, one would specify typical parameter
values of fast cars, slow cars, and trucks, and their respective
percentages. Alternatively, one could introduce a distribution
of each parameter (e.g.~a Gaussian one)
around a typical value. Then, the simulation is started
with the initial and boundary conditions of interest. Of course, the
model can be also extended to a multi-lane model with lane-changing and
overtaking maneuvers \cite{Hab,multilane}. This is a topic of current research.

\subsection*{Acknowledgments}
The authors want to thank for financial support by the BMBF (research
project SANDY, grant No.~13N7092) and by the DFG (Heisenberg scholarship
He 2789/1-1). Moreover, they are grateful to the Bosch GmbH 
for supplying some of their floating car data and to Tilo Schwarz for carrying
out preliminary simulation studies \cite{Tilo}.

\begin{center}
\includegraphics[width=8cm]{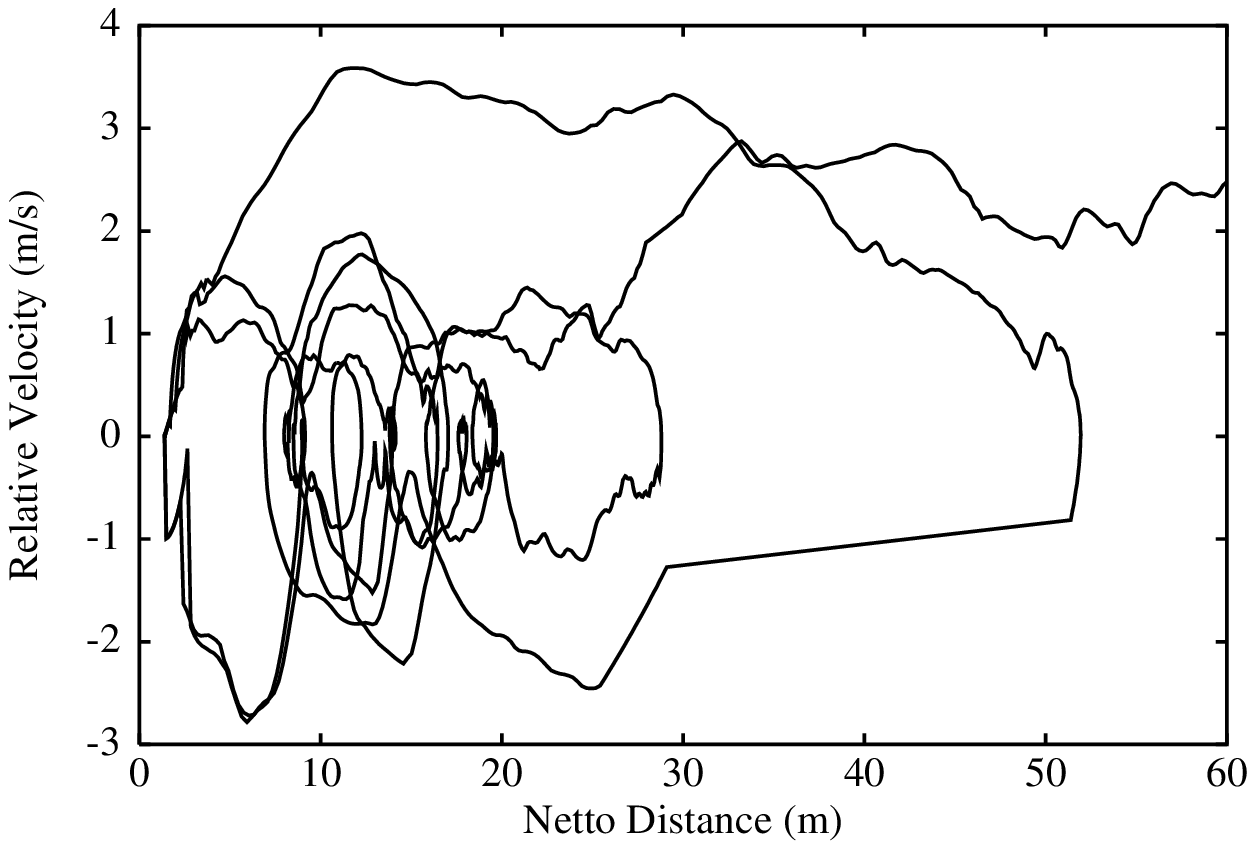}
\end{center}
\begin{figure}
\caption[]{The simulation of the follow-the-leader behavior according
to the generalized force model shows the oscillatory nature of the relative 
motion of vehicles. The above result is in good agreement with the
empirical findings depicted in Fig.~\ref{F6}.\label{F7}}
\end{figure}
\begin{table}[bhtp]
  \caption[]{Minimal values of the average relative deviation $D$ 
  between empirical data and simulation results that were reached
  for the different traffic models by evolutionary parameter 
  optimization.\label{table1} }
  \begin{tabular}{lrrr}
     Model & OVM & T3 & GFM \\
    \hline
    $D$ & 0.0586 & 0.0354 & 0.0316 \\
  \end{tabular}
\end{table}

\end{document}